# *fairness* in society

Ophir Flomenbom

*Flomenbom-BPS, 19 Louis Marshal St., Tel Aviv, Israel 62668*

**Abstract** Models that explain the economical and political realities of nowadays societies should help all the world's citizens. Yet, the last four years showed that the current models are missing. Here we develop a dynamical society-deciders model showing that the long lasting economical stress can be solved when increasing *fairness* in nations. *fairness* is computed for each nation using indicators from economy and politics. Rather than austerity versus spending, the dynamical model suggests that solving crises in western societies is possible with regulations that reduce the stability of the deciders, while shifting wealth in the direction of the people. This shall increase the dynamics among socio-economic classes, further increasing *fairness*.

Economics, sociology, and politics, are all related with the structure of and opinions in the society [1-8]. Models for the dynamics in opinions are vast and diverse and are closely related with problems of interacting entities in physics, chemistry and biology [1, 9-11]. The most encountered model in opinions is the voter model and its variants, e.g. [7-8]. Such models together with theories in sociology and economics should help realizing the behaviors in the society, and should ultimately help organizing fair societies. Yet, since 2007 the world, and in particular western democracies, are facing a financial crisis showing that our understanding is lacking. Here, we develop a new model explaining nations: a system that is composed of those making the decisions, "the deciders", that interact with the bulk of people forming most of the society. One can identify the deciders with the top one in a thousand (0.1%) of the individuals in the nation: business men, politicians, intellectuals, professors, famous ones, and the "clan's elders" including, seniors and formal retiree from the academia, military, politics, and so on. The deciders influence through formal channels, e.g. high rank positions in the public sectors, politics, academia, companies, media, army, etc., and through informal channels, including, accumulated money, connections, family ties, fame, advanced technology of communication, etc. The society-deciders model is explicitly based on the common wisdom that the deciders in the society are in fact a distinct society that controls most of the resources of the country. Our aims here are describing modern societies accurately, and within the model suggest ways of solving economical crises while making a fair society from this unfair realization of the society.

***The model.-*** In the model, the deciders try winning influence in their camp, and surviving. They can hold an unpopular opinion for a long time, risking elimination from the system. The society is characterized with the temperature, $T(t)$, $T(t) = w(t) + \xi(t)$. $w(t)$ is the opinion of the deciders, defined mathematically in *Results*. $\xi(t)$ is a simple Brownian motion that stands for all occurrences in the society and in other societies that affect the system we examine. $\xi(t)$ obeys a simple stochastic equation of motion: $\dot{\xi} = c\sigma$, where $\sigma(t)$ is a Gaussian noise with a zero mean and a unit variance, where $c$ determines the strength of the noise. Clearly, $T$ can have any real value. Note that $c$ has the units of the temperature and $dt\sigma(t)$ is a quantity without units.

***The deciders.-*** The model for the deciders is an equation of motion for $X_\delta(x,t)$: the number of deciders with a particular level of influence $x$ at time $t$ with opinion $\delta$, where $\delta = +, -$. The dynamics of $X_\delta(x,t)$ follows,

$$\partial_t X_\delta = (D\partial_{xx} + v\partial_x)X_\delta + k_{b,\delta}\delta(x)$$

$$-\big(k_{d,\delta}(x) + f_\delta(x)\big)X_\delta - (c_\delta - c_{-\delta})X_\delta \tilde{X}_{-\delta} \qquad ; \qquad \delta = +, -. \quad (1)$$

The deciders diffuse in the influence coordinate $x$, where a constant force that attracts towards the origin is applied, since it is hard achieving influence in the system. At the origin there is a reflecting boundary. Birth and death terms are included. Deciders join the system at the origin. The death term represents the death of deciders with influence $x$. The death and the birth terms are balanced most of the time, and only in extreme conditions a net change of deciders is observed. Terms for changes in the opinions are included: A stress term represents those that decided starting from scratch with a different opinion. A reaction term represents those that decided switching groups after an encounter with deciders of at least a similar influence yet a different opinion. In this situation, the decider switches opinion yet manages having the exact level of influence obtained during the years in the other group. Note that all the dynamical functions that control the dynamics and survival of the deciders depends on $e^{\delta \frac{T}{\tilde{T}}}$, making the connection among the deciders and the people. $\tilde{T}$ is therefore the interaction strength among the society and the deciders.

***The rate functions in the model.-*** The death function $k_{d,\delta}(x)$ follows:

$$k_{d,\delta}(x) = k_{d,\delta}(0)e^{-\gamma_{d,\delta}\frac{\tilde{x}}{x_0}} \quad ; \qquad \gamma_{d,\delta} \geq 0. \tag{2}$$

Here, $\tilde{x} = |x - x_{max}|$, where $x_{max}$ is the value of $x$ that maximizes $X_\delta(x,t)$. When $\gamma_{d,\delta} > 0$, Eq. (2) means that those with a large influence are relatively more stable, where those with little influence win grace and are also relatively more stable comparing with those at $X_\delta(\tilde{x} = 0, t)$. A stress term $f_\delta(x)X_\delta$ represents those that decided starting from scratch with a different opinion, where,

$$f_\delta(x) = f_\delta(0)e^{-\gamma_{f,\delta}\frac{\tilde{x}}{x_0}} \quad ; \qquad \gamma_{f,\delta} > \gamma_{d,\delta}. \tag{3}$$

This term decays faster with $x$ relative with eq. (2) since it is assumed that the basic tendency of those with a lot of influence is staying in their camp. The coefficient $f_\delta(0)$ may fight this tendency while introducing the influence of the temperature on the activity of the deciders. This point is presented in what follows. Note that in eq. (3), we use $\tilde{x}$ in accordance with eq. (2).

All the coefficients depend on the temperature. The fluctuations rates follow: $c_\pm \rightarrow c_\pm\, e^{\pm 2\frac{T}{\tilde{T}}}$, where the death rates follow: $\tilde{k}_{d,\pm} \rightarrow \tilde{k}_{d,\pm}\, e^{\pm\frac{T}{\tilde{T}}}$. The fear function's coefficients $f_\pm(0)$ obey the equation: $f_\pm(0) \rightarrow f_\pm(0)e^{\pm a\frac{T}{\tilde{T}}}$. These relations introduce economic relations in the model, since the deciders depend on the temperature. Possible generalizations are considered in the supplementary information (SI) part A.

***The coefficients in the model.-*** Nine of the coefficients in the model are determined from physical considerations. Five others are tunable (the powers, $\gamma_{f,\delta}, \gamma_{d,\delta}, a$), yet always coincide with the general behavior found here (obtained when all these powers are one). The external coefficients are the noise strength $c$ and the temperature of stability, $\tilde{T}$.

Firstly, we let $D = 1$, simply scaling the time. Then, we also scale the 'energy', while letting, $v = 1$. Finally, we scale the distance with, $x_0 = 1$. We relate the death rate and the birth rate when balancing the death with the birth: $k_{b,\delta}\delta(x) = k_{d,\delta}(x)X_\delta \rightarrow k_{b,\delta} = \int k_{d,\delta}(y)X_\delta(y,t)dy$, making the birth rate a time dependent coefficient. Still, when the temperature is large (in absolute value), this relation does not necessarily hold, since we set the maximal value of the death rate (at any opinion) with $n_\delta(0)/20$ individuals per month ($n_\delta(0)$ is the number of deciders with opinion $\delta$ at the initial stage of the dynamics, and the maximal birth rate with $n_\delta(0)/100$ per month.

Now, the following values for the death and fluctuations rates are considered: $k_{d,+} = 35year^{-1}$ and $k_{d,-} = 35year^{-1}$, $c_+ = 20 \ years^{-1}$ and $c_- \approx c_+$. It is assumed that a decider enters the system at the age of about 35, so the average lifetime and standard deviation of an individual in the system is $70 \ years \pm 35years$, and there is a reasonable chance of changing an opinion once with saving one's power. Note also that the dependence on the temperature introduces inequality in the relations among the rates of different opinions. Finally, $f_\delta(0) \approx 7year^{-1}$, meaning that those with little power may switch groups about once in a decade.

**RESULTS.-**

***Numerical results.-*** We record the number of deciders in each of the opinions, $n_\delta(t)$, and $T(t)$, and $w(t)$. $w(t)$ is defined with, $w(t) = c\frac{\sum_x x(X_+(x,t)-X_-(x,t))}{\sum_x[X_+(x,t)+X_-(x,t)]}$. The coefficient $c$ is used in $w(t)$ for balancing the noise. The results are collected in the area $c\tilde{T}$. Figure 1 shows typical trajectories of $w(t)$, $T(t)$ , left panels, and $n_\pm(t)$, right panels. The trajectories of $T(t)$ reflects the situation in society. Large deviations from the origin are bad times. The deciders can ignore the people (small and intermediate $c$ with large $\tilde{T}$, panel 1A), or show sensitivity to their condition (small and intermediate $c$ with intermediate $\tilde{T}$, panel 1B). When $c$ is very large and $\tilde{T}$ is even larger, the people can collapse (Panel 1C).

***Stability.-*** Stability in the deciders is when $n_{\pm}(t)$ fluctuate around the initial values. Collapse happens when a particular deciders' camp does not survive and has about 10% of the initial population, or when the number of deciders in a particular camp is about 10 times larger than in the initial stage. Small fluctuations in $T(t)$ around the origin are a sign of stability in the society. The occurrence of large time interval where $|T(t)|$ is large indicates that the society is in a crisis. This is defined when $|T(t)|$ reaches five times the amplitude of the oscillations in $|T|$ in the region of stability.

We collect statistics of the deciders' collapse. The main conclusion shows a rather straight line, $\tilde{T}(c) \sim c$, in the area $c\tilde{T}$ that defines a region where the deciders do not collapse and a region where they do collapse. In the region where deciders collapse, the temperature is usually stable, and $\tilde{T}$ is relatively small indicating on intermediate-to-strong interactions. When the interactions are considered small (large $\tilde{T}$), the system is usually stable for small-to-intermediate noise and only when the noise increases a lot, the system can collapse due to the temperature. Numerical results are presented in the SI part B.

***Mathematical results.-*** We focus on the dynamics among the opinions, and study the stability in $X_\delta$ in a simplified version of the society-deciders model, showing that the relation $\tilde{T}(c) \approx c$ is indeed the border of stability. See the SI part B for further information.

**Discussion.-**

***The relative noise.-*** The society-deciders model has a stable mode and a mode that can collapse depending on $c/\tilde{T}$, the relative noise. Collapsing represents a situation where the top one in a thousand needs the help of the next group in the society for maintaining control. Only when the deciders can collapse since the conditions of the people are extreme, fairness in the society is achievable. This happens when the relative noise is about one. We relate the relative noise with occurrences and characterizations of the society, while writing,

$$\frac{c}{\bar{T}} = \frac{c_1}{\bar{T}_1} e^{\Sigma_i \tilde{u}_i - \Sigma_i \tilde{q}_i}, \tag{4}$$

where,

$$c = c_1 e^{\tilde{u}_{individual} + \tilde{u}_{nation} + \tilde{u}_{global}} \qquad ; \qquad \tilde{u}_i = u_{freedom}\left(\frac{u_i}{<u_i>} - 1\right), \tag{5}$$

and,

$$\bar{T} = \bar{T}_1 e^{\tilde{q}_{individual} + \tilde{q}_{nation} + \tilde{q}_{global}} \qquad ; \qquad \tilde{q}_i = q_{freedom}\left(\frac{q_i}{<q_i>} - 1\right). \tag{6}$$

All events in the $\tilde{u}_i$s and the $\tilde{q}_i$s are indicators in economics and politics, most of these change on the scales of months and years, namely, characterizes society in the thermodynamic limit of $n, t \to \infty$ (where $n, t$ represent the number of individuals and the time, respectively). The most logic function when using the indicators in this limit is an exponent. Note that all indicators used here are mostly in the range zero and one. The rates $u_{freedom}, \tilde{u}_{individual}, \ \tilde{u}_{nation},$ $\tilde{u}_{global}$ are all related with events that affect the general public in the nation, where $q_{freedom}, \tilde{q}_{individual}, \ \tilde{q}_{nation}, \ \tilde{q}_{global}$ are all related with events that affect the deciders of the nation. Each event is part of a category: (1) the category, $individual$, contains every event within the nation that affects or characterizes the stability of the deciders for the $\tilde{q}_i$s and of the people for the $\tilde{u}_i$s. (2) $nation$ contains every rule, regulation and law that affects the stability of the deciders for the $\tilde{q}_i$s and of the people for the $\tilde{u}_i$s. (3) $global$ contains every influence from other nations and organizations that directly affects the stability of the deciders for the $\tilde{q}_i$s and of the people for the $\tilde{u}_i$s.

In eqs. (4)-(6), every one of the rates can have several contributions; for example,

$$\tilde{u}_i = \mu_{u,i,1}\tilde{u}_{i,1} + \mu_{u,i,2}\tilde{u}_{i,2} + \mu_{u,i,3}\tilde{u}_{i,3} + \cdots, \qquad\qquad (7)$$

where $\mu_{u,i,j}$ is the strength of rate $j$ in $\tilde{u}_i$. When additional information is not known, the $\mu_{ij}$s are all equal for any particular $i$. The $\mu_{u,i,j}$s are normalized: $\Sigma_{i,j}\mu_{u,i,j} = \mu_{u,total}$, where $\mu_{u,total}$ is user defined; e.g., $\mu_{u,total} = 3, 9, \dots$ This normalization introduces another degree of freedom in the computations. It also means that we can use the largest contribution among all events like an upper bound for the total, e.g., $\Sigma_i\tilde{u}_i \leq \mu_{u,total}\max_{i,j}\{\tilde{u}_{i,j}\}$. We use this relation in calculations of new events. Yet, in the equations for the $c_1$ and $\tilde{T}_1$ we choose the events that change relatively slowly and the coefficients are chosen such that these will not create user-defined-bias in the result. Clearly, once a set of coefficients is determined it is applied on all the nations.

***Several notes:*** 1) we can incorporate the magnitude of the event in the computations. 2) $\mu_{ij}$ can introduce effects that are not linear in the variable $category$. 3) $\tilde{u}_{category}$ can have a negative value indicating that the category decreases the noise, where negative value for $\tilde{q}_{category}$ decreases the temperature of stability.

A specific discussion about the computations of the rates is presented in the SI part C. Here, we introduce an example, while computing the rate of the unemployment in the USA, $u_{individual,unemploy.}$ (where we used in most cases simply the notation, $u_{unemployment}$). We use the unemployment rate per month that is published at http://www.bls.gov/, $u_{unemployment} = \frac{unemployment\ rate}{100\%}$, so $u_{unemployment}$ is a number in the range zero and one. For computing the average rate, e.g. $< u_{i,j} >$, we can simply perform an average over a period of a year when the event is computed per month, or average over five years when the event in computed every year. Yet, we can set the average rate like the rate that is considered the normal one, or the average one, or the desirable one, or the most relevant one, etc. For example, $< u_{unemployment} > = 5\%$, calculated as the mean in the period 2001-2006.

**Computing $\tilde{T}$:** $\tilde{T}$ determines the interaction's strength among the deciders and the people. It thus introduces the temperature in the system of the deciders. $\tilde{T}$ includes a simple combination of the following indicators: the relative wealth of the top one percent, $q_{wealth}$, the tradition in the nation, $q_{tradition}$, the democracy index [12], $u_{democracy}$, the strength of the nation in the world [13], $q_{strength}$, and its IMF's shares [14], $q_{IMF}$. In particular, $\tilde{T}_1$ follows, $\frac{ln(\tilde{T}_1)}{q_{freedom}} = q_{individual} + q_{nation} + q_{Global}$, where the indicators that are used in computing the three categories in $\tilde{T}_1$ include: $q_{individual} = \frac{1}{2}(q_{wealth} + q_{tradition})$, $q_{nation} = \frac{4}{3}u_{democracy}$, and $q_{global} = \frac{1}{2}(q_{IMF} + q_{strength})$. Note that the expressions of both $c$ and $\tilde{T}$ depend on the freedom index [17].

**Computing $c$:** the noise influencing the people includes a simple combination of the following indicators: the human development index [15], $u_{HDI}$, the income spread [16] (the Gini index), $u_{Gini\ index}$, the democracy index, $u_{democracy}$, the nation's debt, $u_{debt}$, and gross domestic product (GDP), $u_{GDP}$, and the nation's conflicts, $u_{con.}$, diplomatic ties, $u_{diplomatic\ ties}$, and trading relations, $u_{diplomatic\ ties}$. In particular, $c_1$ follows, $\frac{ln(c_1)}{u_{freedom}} = u_{individual} + u_{nation} + u_{Global}$, where the indicators that are used in computing the three categories in $c_1$ include: $u_{individual} = \frac{1}{2}(u_{Gini\ index} + u_{HDI})$, $u_{Global} = \frac{1}{3}(u_{economical\ ties} + u_{con.} + u_{diplomatic\ ties})$, and $u_{nation} = \frac{1}{3}(u_{democracy} + u_{debt} + u_{GDP})$.

**Applying the model.-** When using the society-deciders model, we simply need calculating the relative noise, eq. (4), from the available data on the particular nation. We apply the model on various nations while considering events in 2009 through 2011. We calculate $c_1$, $\tilde{T}_1$, $R = \frac{c}{\tilde{T}} / \frac{c_1}{\tilde{T}_1}$, and *fairness*, where,

$$fairness = \tilde{T}_{1,ideal} / \tilde{T}_1, \qquad\qquad (8)$$

and $\tilde{T}_{1,ideal}$ is $\tilde{T}_1$ in an ideal nation, a number of the order unity. We also suggest best ways of improving. Results are presented in table 1, where information on the calculations on every nation is presented in the SI part C.

***The conclusions:***

- ***The main conclusions.-*** Indeed, $c$ increases in economical crises. While in a crisis, nations try reducing $c$, where $\tilde{T}$ is usually large and unchanged. Within the model, solving the profound economic problems rather than simply buying time until the next crisis requires putting in place measures that reduce $\tilde{T}$, namely, increasing *fairness* in society. Including new tax steps applied on the top 1% with new governmental programs in favor of the people is the simplest way of shifting wealth in the direction of the people. Such measures will reduce the wealth of the one in a thousand, and will reduce the gaps among all classes in society, yet also improve the dynamics among classes, namely, will help increase the opportunities in the entire society. Note that in the particular case of the USA, the analysis suggests that for solving the economic crisis in the USA while reducing $\tilde{T}$, public awakening is the best first step. Elaboration is presented in the next points.

- ***The basic conclusions.-*** 1) The most basic thing in a nation is increasing the freedom and democracy of its citizen. Then, we look on the local and global economical and political indicators and compute the fairness in the nation. The tradition in a nation influences on the stability of the deciders. Education can reduce this effect. 2) Lasting mass protests signal that the deciders collapsed. 3) The noise coefficient increases in times of a crisis. Although reaching a large value in $|T|$ has also random contributions, the probability of reaching a large value increases when $c$ is larger for a constant $\tilde{T}$. This means that in a crisis, we will compute, most probably, a relatively large $c$. The debt, GDP, unemployment and poverty rate are all indicators that in a time of economical crisis increase $c$. The temperature of stability $\tilde{T}$ is unchanged when these indicators increase. Protests also increase $c$, where $\tilde{T}$ is unchanged. Yet, protests give a serious additional motivation for the deciders "to wake up" and solve the crisis. 4)  Increasing or decreasing the debt of the nation is not the most important debate within the people-deciders model when trying solving the crisis; the effect of any new regulation on the stability of the deciders is the crucial issue.

- ***Redistribution of the wealth.-*** The most important way of solving an economical crisis in a nation, while increasing fairness in society, is decreasing the wealth of the top one percent. (Mathematically, we see this from Eqs. (C7) through (C14) in the SI part 3.) Note that since the society is built like an onion of socio economic classes, this means that the entire top 1% will lose power in favor of the others. For seeing this, we label the socio economic layers with index $i$, where $i = 1$ is for the top one in a thousand, $i = 2$ is for the next 0.9%, $i = 3$ represent the next 9%, $i = 4$ represents the next 10%, and $i = 5$ represents the next 80%. With this notation, the current allocation of wealth in the USA follows, e.g. [18]:

$wealth = \{17.5\%, 17.5\%, 35\%, 15\%, 15\%\}$. In a fair society, we suggest the following:
$wealth = \{4.5\%, 7\%, 38\%, 18\%, 32.5\% \}$. Note that any class improves its condition relative with the next layer in the new situation, so the one in a thousand lose power relative with the entire society. Note that class 5, the poorer class, contains at least eight mini classes.

   This suggestion for a new redistribution in wealth is possible with additional tax rates for those making more than 350,000\$, in the USA. The taxation should continue even until a tax rate of 75%, affecting also billionaires. Indeed, there are other possibilities for redistribution (e.g., see the discussion around eq. (C8) in the SI part C). In any suggestion of redistribution of the wealth, the clear thing is that the top 1% loses a large portion of its wealth, and the conditions of the other 99% of the people improve significantly. Clearly, the money will reach the people with new governmental programs in housing, education, health care, etc.

- **Power and control.-** Indeed, the wealth also measures power and control in the society. In a healthy society, rather than speaking on the top one in a thousand, we should have several percentages at the top (say three) that are equally important in maintaining power and control. Since the wealth distribution is highly distorted also at its peak, it is obvious that the one in a thousand is in control, and the next class in the society's onion actually protects the one in a thousand. Thus, in terms of power and control, a healthier society has the following allocation of groups and power,

$power = \{[18\%]_{97\%}, [13\%]_{93\%}, [9.67\%]_{87.67\%}, [7.67\%, ]_{80\%,}, [53\%]_{0\%} \}$. Here, the number inside the brackets represents the total power of the group ending at the percentile indicated with the subscript value.

- **About the dynamics.-** The dynamics among the socio economical classes in the society's onion should increase in a fairer society. Reducing the wealth and power of the one in a thousand should enable faster dynamics among classes. We can quantify this when computing the average wealth per individual in each class of the society's onion and the relative wealth per individual among layers. This simple analysis shows that the wealth per person in the various classes in a healthier nation follows: $wealth\ per\ person = \{65, 9.8, 2.37, 1, 2/3\}$, so that the ratios among adjacent classes of the wealth per individual, $W$, follows: $W = \{6.57, 4.17, 2.37, 1.5\}$. The ratios should indicate on the distance in wealth (power) among classes. Note that in the current situation in the USA, $W$ follows: $W = \{9, 4.86, 2.59, 8\}$. Thus, when redistributing the wealth in the society, the distance among classes decreases. We suggest that the rate of moving among classes depends on the distance, and this means that the dynamics are faster. For example, the rate of moving from $i = 2$ and reaching $i = 1$ follows: $k_{2\to1} = k_{2\to1}^0 e^{-W(1)}$. Since $W(i)$ in a fairer society is always smaller, the dynamics are faster. The nation should further increase the zeroth rate of moving among classes, e.g. $k_{2\to1}^0$, with new rules and regulations.

The life time of individuals in a class follows: $\tau_i = \frac{1}{k_{i\to i+1} + k_{i\to i-1}}$. ($k_{1\to0} = k_{5\to6} = 0$). We can say that in a fair society $\tau_1 \approx 5\ years$. In addition, $\tau_2 \approx 5\ years$, and all other lifetimes are about 10 classes, where the poorer class is treated in mini classes of 10%, where all these mini classes are connected with the forth class (with different rates).

### *References*

[1] Smith A., *The Theory of Moral Sentiments* (Edinburgh) 1759.

[2] Varian H. R., *Intermediate Microeconomics: A Modern Approach* (W.W. Norton & Co., NY, NY; Eighth Edition) 2010.

[3] Krugman P. R., Obstfeld M., *International Economics: Theory and Policy* (Addison Wesley, Boston, MA, USA; 6 edition) 2002.

[4] Schelling T. C., *J. Math. Soc.*, **1** (1971) 143.

[5] Colander D., Goldberg M., Haas A., Juselius K., Kirman A., Lux T., Sloth B., *Cri. Rev.*, **21** (2009) 249

[6] Lemoy R., Bertin E., Jensen P., *EPL* **93** (2011) 38002.

[7] Liggett T. M., *Stochastic interacting systems: contact, voter, and exclusion processes* (Springer, Germany-Verlag Berlin Heidelberg) 1999.

[8] Sood V., Redner S., *Phys. Rev. Lett.*, **94** (2005) 178701.

[9] Flomenbom O., Taloni A., *Europhys. Lett.*, **83** (2008) 20004.

[10]. Flomenbom O., *EPL*, **94** (2011) 58001.

[11] Flomenbom O., *Phys. Lett. A*, **374** (2010) 4331.

[12] The report about the democracy index:

http://www.economist.com/media/pdf/DEMOCRACY_INDEX_2007_v3.pdf

[13] Chapnick A., *Canadian Foreign Policy*, **7** (1999) 73.

[14] The IMF shares of all nations are presented at,

http://www.imf.org/external/np/sec/memdir/members.aspx

[15] The report about the human development index:

http://hdr.undp.org/en/media/HDR_2010_EN_TechNotes_reprint.pdf

[16] Gini C., Concentration and dependency ratios (in Italian) 1909. English translation in *Rivista di Politica Economica*, **87** (1997) 769.

[17] The report about the Freedom Index:

http://www.freedomhouse.org/template.cfm?page=351&ana_page=373&year=2011

[18] William Domhoff G., *Who Rules America? Challenges to Corporate and Class Dominance*

(McGraw-Hill Humanities, N.Y.) 2009.

**Figure captions**

**Fig1.** (color online) Numerical results of the model. The deciders are distributed in an interval of length $L$ like, $X_\delta(x, t \to 0) = 2 \frac{n_\delta(0)}{L} (1 - \frac{x}{L})$, where in the simulations, $n_\delta(0) = 10{,}000 \; individuals$, $L = 55$, $dx = 0.3$ and $dt = 0.03 \; months$, with $t_{max} = 94 \; years$. Here we show the number of individuals in each of the camps (right, $n_+$ in azure), and the opinion of the deciders and the temperature (left, $w$ in green). $\tilde{T} = 0.177$ & $c = 0.377, 3$, in the first two lines, and $\tilde{T} = 2.577$ & $c = 3$ in the third line.

**Table captions**

Table 1. A concise representation of the results for 18 nations. These nations appeared constantly in the world wide media around the world in the several last years.

**Table 1**

| Country | $fairness$ | $R$ | Solutions for improving fairness |
|---|---|---|---|
| China | 0.01 | $R \leq 1$: protests & oppression of protests | Move in the direction of democracy & freedom. Improve the wealth distribution. |
| Russia | 0.0065 | $R \leq 1$: protests & oppression of protests | Improve democracy & freedom. Decrease the relative wealth of the top 0.1%. Improve the wealth distribution. |
| Japan | 0.306 | $R \geq 1$: tsunami and government shuffle | Decrease the relative wealth of the top 0.1%. |
| Spain | 0.58 | $R \geq 1$: mass protests, high unemployment | The austerity measures should decrease the relative wealth of the top 0.1%. Decrease the unemployment rate. |
| Greece | 0.51 | $R \geq 1$: mass protests, debt problem | The austerity measures should decrease the relative wealth of the top 0.1%. |
| UK | 0.447 | $R \geq 1$: protests & economic stress | The austerity measures should decrease the relative wealth of the top 0.1%. |
| France | 0.438 | $R \approx 1$: Recent economic stress are balanced with new regulations in France | Decrease the relative wealth of the top 0.1%. |
| Belgium | 0.58 | $R \geq 1$: political stress | Decrease the relative wealth of the top 0.1%. Stabilize the political system |
| Germany | 0.67 | $R \geq 1$: Local economical stress, yet, deciders won points in improving the EU economical stress. | Decrease the relative wealth of the top 0.1%. |
| Sweden | 0.84 | $R \approx 1$. | Decrease the relative wealth of the top 0.1% |
| Egypt | 0.154 | $R \gg 1$: revolution of the people; shift in the direction of democracy | Establish a stable democracy |
| Libya | 0.142 | $R \gg 1$: revolution of the people; shift in the direction of democracy | Establish a stable democracy |
| Syria | 0.0044 | $R \to 0$: revolution of the people; violent oppression of protests | Change the regime while forming a democracy |
| Saud Arabia | 0.0032 | $R > 1$: protests; new regulations | Form a democratic and free society |
| Iran | 0.002 | $R \ll 1$: mass protests; violent oppression of protests | Change the regime while forming a true democracy |
| Israel | 0.4 | $R \leq 1$: mass protests; stagnation with the peace agreement, economical stress | Decrease the relative wealth of the top 0.1%. Improve the wealth distribution. Make a peace agreement |
| USA | 0.2 | $R \leq 1$: long economical stress; the political stagnation is apparent, and the gap among the top one percent and the others increased in recent years | Decrease the relative wealth of the top 0.1%. Improve the wealth distribution. Decrease the blind appreciation for money and fame. |
| Brazil | 0.135 | $R \approx 1$. | Decrease the relative wealth of the top 0.1%. Improve the wealth distribution. |

**Figures**

FIGURE 1

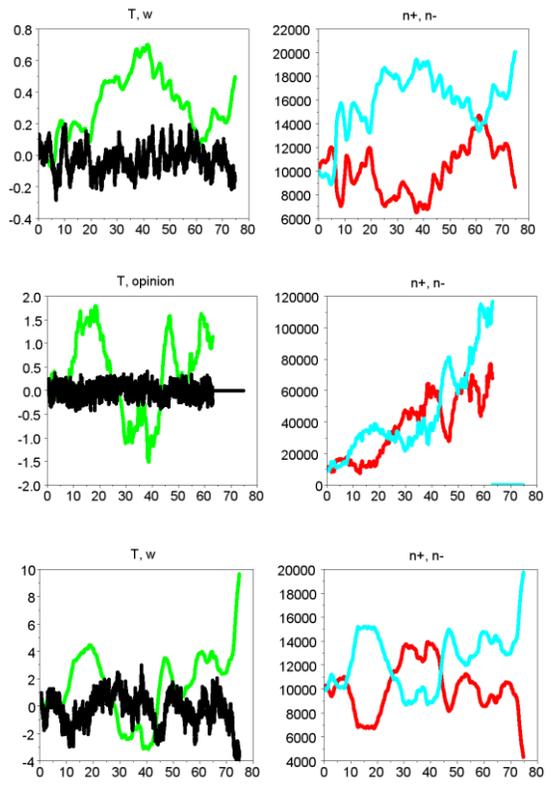

# Appendices: *fairness* in society

Ophir Flomenbom

*Flomenbom-BPS, 19 Louis Marshal St., Tel Aviv, Israel 62668*

**Appendix A: *Extensions in the model.-***

In this appendix we consider several possible extensions in the basic models. The generalizations may include:

- The diffusion operator $L(x)$ can have a form that is more complicated in the coordinate $x$, or terms that are not linear in the function $X_\delta$. Such effects are not considered here, since we considered the simplest and most reasonable form for a potential reflecting the fact that there is an exponential hierarchy in the deciders society.

- $c_\delta$ can have a dependency on $x$, $c_\delta(x)$. This is also not considered here since the term $c_\delta X_\delta \tilde{X}_{-\delta}$ is complicated in its current form, and small for large $x$ since it is quadratic in $X_\delta$ that is exponentially small in the coordinate $x$ at large $x$.

- In a complicated variant of this model, another coordinate is included; this represents, for example, the level of accumulated achievements of the decider. Here, the influence $x$ represents all the benefits in the model.

- A continuous opinion coordinate can also be considered. Here, the combination of a discrete opinion coordinate and a continuous influence coordinate is used instead.

- Other functional form for $f_\delta(x)$ and $k_{d,\delta}(x)$ can also be considered, yet we choose here the most physical ones in this model; since the steady state solution of $X_\delta$ is exponential, the exponential forms for $f_\delta(x)$ and $k_{d,\delta}(x)$ make these terms comparable with the reaction term.

- Other functional forms of the rates dependency on the temperature are possible. Still, we choose here the most logical form: exponentials make these terms similar in magnitude, and all terms in the equation of motion for $X_\delta$ can compete each other.

- Introducing extremists with influence $x \geq X$, these folks do not change their opinion but simply die in extreme conditions. We do not study here this variant of the model.

- Finally, we note that the model can even help solving conflicts and disputes among nations, like the conflict between Israel and Palestine authority. We will consider generalizations involving variants with different nations in a forthcoming publication.

- Illustration of the model is shown in fig. A.1.

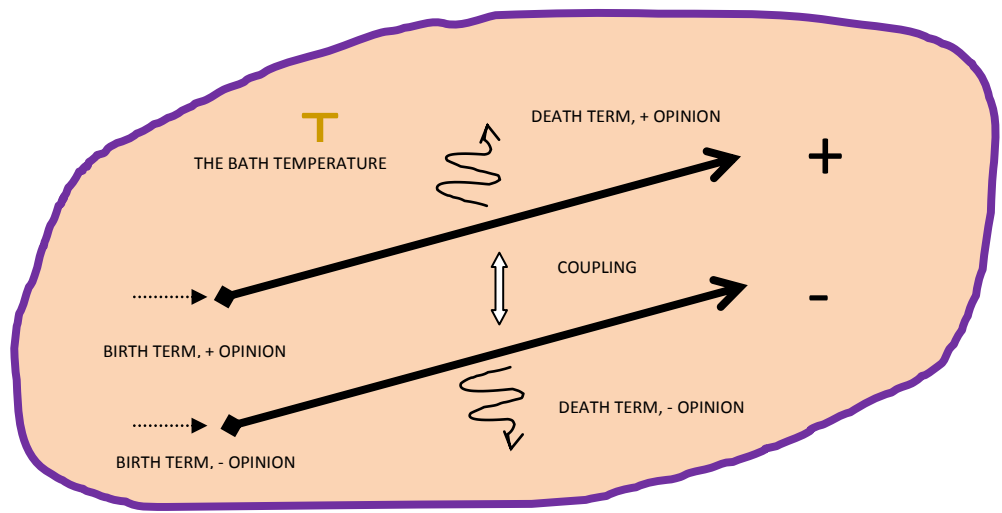

**Figure A.1.** An Illustration of the Model.

***Appendix B: Supplementary numerical and mathematical results***

In this model, we measure the number of deciders in each camp, $n_\delta(t) = \Sigma_x X_\delta(x,t)$, the temperature $T(t)$ and the total opinion $w(t)$, while changing various coefficients in the model.

We look on the stability in the system: the stability is a function of the number of deciders in both camps relative with their initial number and the value of the temperature.

Recall that we find that a simple straight line in the area spanned with the coordinates $\tilde{T}$ and

$c$ defines a region where the deciders collapse and a region where the deciders do not collapse.

Here, $c$ is the noise strength and $\tilde{T}$ determines the strength of the interactions of the deciders and the people, and is seen like the temperature of stability, where small $\tilde{T}$ represents a large interaction's strength. It is the aim in this appendix presenting additional results regarding the behaviors in the model while changing the values of several coefficients and presenting mathematical results for a simplified model of the original one.

***B1. Numerical results in the area*** $c\tilde{T}$**.-** We start this part with additional results regarding the numerical behavior in the model. We present results that are complementary with those shown in Fig. 1 in the main text, where all the coefficients and constants match: $a = 1$, $\gamma_{d,\delta} = 1$, and

$\gamma_{f,\delta} = 1$ and $\gamma_w = 1$. Results are for $w(t)$, $T(t)$ , left panels, and $n_\pm(t)$, right panels. The results

further support the behavior explained in the main text about these quantities.

In the first line in fig. B.1, results were obtained for, $\tilde{T} = 3.77$, and $c = 1$. These panels show that when the noise strength $c$ is small and the temperature of stability $\tilde{T}$ is very large, the society and the deciders are independent. This may cause the society reaching a very high temperature in absolute value, where the time reaching these values depend on the $c$. This behavior is enhanced when increasing $\tilde{T}$. See the second line in fig. B.2, where , $\tilde{T} = 38$, and

$c = 1$. This behavior is also enhanced when increasing $c$ when $\tilde{T}$ is very large. See the third line in fig. B.2, where , $\tilde{T} = 38$, and $c = 7$.

For a fixed $\tilde{T}$, increasing $c$ increases the system's volatility until collapsing. For example, setting, $\tilde{T} = 2.77$, and varying $c$, $c = \frac{1}{4}, 13, 38$, we find that the probability for collapse and the time for collapsing follow, respectively, $p_{col} = \frac{1}{3}, \frac{4}{10}, \frac{2}{3}$, and $t_{col} = 49, 47, 54\ years$. The results were computed when averaging over the noise on 27 realizations. Recall that collapse can occur since one camp reaches small values and when one camp reaches high values or when the temperature is very large.

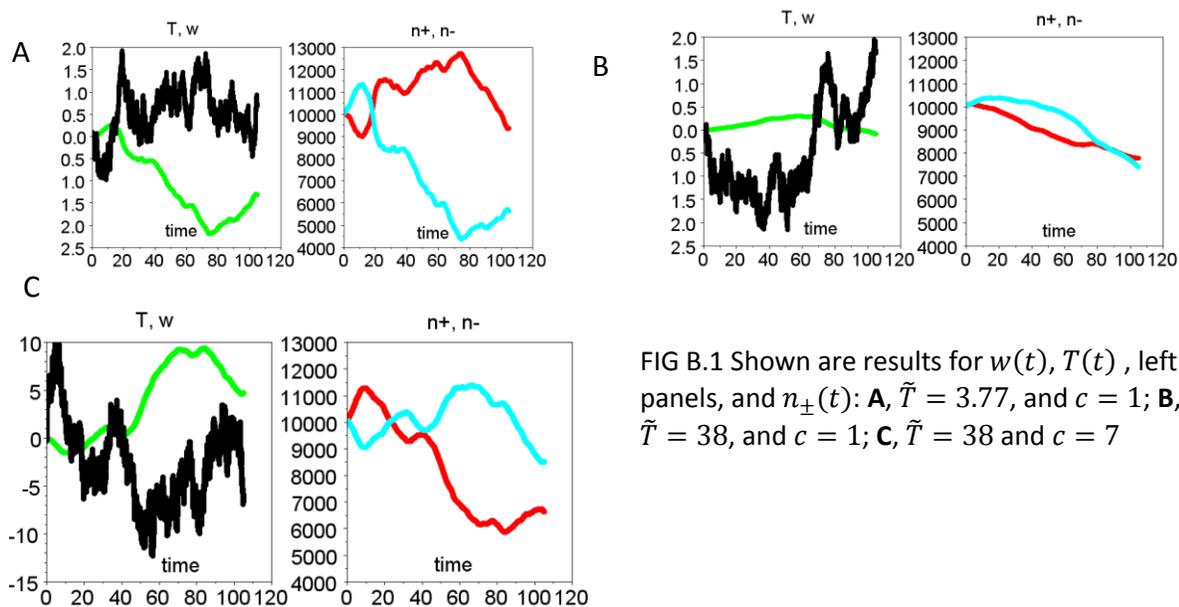

FIG B.1 Shown are results for $w(t)$, $T(t)$ , left panels, and $n_{\pm}(t)$: **A**, $\tilde{T} = 3.77$, and $c = 1$; **B**, $\tilde{T} = 38$, and $c = 1$; **C**, $\tilde{T} = 38$ and $c = 7$

***Collapse statistics.-*** Figure B.2 shows results about the collapse statistics: $p_{col}$ the probability for collapsing, $t_{col}$ the time for collapsing. In this figures, collapses are because deciders collapse.

Like the above, we present results that are complementary with those shown in fig. 1 in the main text, where all the coefficients and constants match: $a = 1$, $\gamma_{d,\delta} = 1$, and $\gamma_{f,\delta} = 1$ and $\gamma_w = 1$.

From the results it is clear that there is a rather straight line, $\tilde{T}(c) \sim c$, in the area $c\tilde{T}$ that defines a region where the deciders do not collapse and a region where they do collapse. In the region where deciders collapse, the temperature is usually stable, and $\tilde{T}$ is relatively small indicating on intermediate-to-strong interactions. When the interactions are considered small (large $\tilde{T}$), the system is usually stable for small-to-intermediate noise and only when the noise increases a lot, the system can collapse due to the temperature.

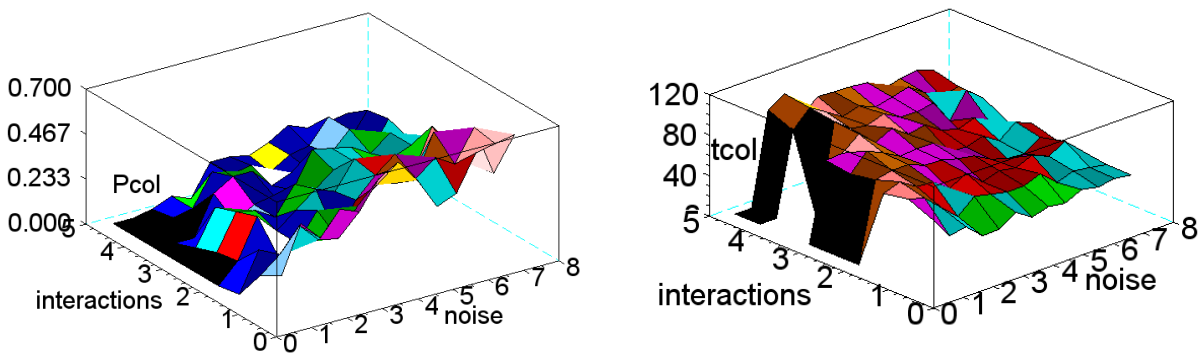

**Fig 2.** Results in the area $c\tilde{T}$ of several quantities describing collapse. The results are collected in the area $c\tilde{T}$ when averaging over 49 realizations. The collapse probability increases in the area times $\tilde{T} \leq c$ where the time for collapsing reduces in this area.

***B2. Numerical results for the various powers in the area $c\tilde{T}$.-*** Here, we present results while changing the values of the powers $a$, $\gamma_{d,\delta}$, and $\gamma_{f,\delta}$. For each value of $a$, $\gamma_{d,\delta}$, and $\gamma_{f,\delta}$, results are for $\tilde{T} = 3.77$, and $c = 1/4$, 3,13.8.

- $a = 1$, $\gamma_{d,\delta} = 1$, and $\gamma_{f,\delta} = 1$. This is the reference, with the results: $p_{col} = 0.55, 0.518, 0.59$ and $t_{col} = 65, 54, 44\ years$.

- $a = 1$, $\gamma_{d,\delta} = 3.77$, and $\gamma_{f,\delta} = 3.77$. The temperature effect is unchanged, where other effects decrease. We find an increased stability: $p_{col} = 0.55, 0.4, 0.518$ and $t_{col} = 51, 70, 52\ years$.

- $a = 1$, $\gamma_{d,\delta} = 0.113$, and $\gamma_{f,\delta} = 0.113$. The temperature effect is unchanged, where other effects are increased. We find a decreased stability: $p_{col} = 0.7, 1, 1$ and $t_{col} = 52, 57, 45\ years$. Note that most of the effect result from deceasing $\gamma_{f,\delta}$: when $a = 1$, $\gamma_{d,\delta} = 1$, and $\gamma_{f,\delta} = 0.113$, we see, $p_{col} = 2/3, 1, 1$, where when $a = 1$, $\gamma_{d,\delta} = 0.113$, and $\gamma_{f,\delta} = 1$, we see, $p_{col} = 0.62, 2/3, 3/4$. The reason is that conservation of the deciders with regard of death and birth exists in this model most of the time, where when the deciders change opinions and start over faster they create a gap since the system losses influence and its capability of fighting fluctuations in the temperature, and therefore the deciders are more fragile.

- $a = 3.92$, $\gamma_{d,\delta} = 3.77$, and $\gamma_{f,\delta} = 3.77$. The temperature effect in the starting over term is enhanced, where other effects are decreased. We find an increased stability for weak and intermediate noise and a decreased stability at large noise: $p_{col} = 0, 0.37, 1$ and $t_{col} = --, 33, 9.45\ years$.

- $a = 0.33$, $\gamma_{d,\delta} = 0.113$, and $\gamma_{f,\delta} = 0.113$. The temperature effect in the starting over term is decreased, where other effects are enhanced. We find an increased stability for weak noise and decreased stability at intermediate and large noise: $p_{col} = 0.14, 0.96, 1$ and $t_{col} = 103, 69, 59\ years$.

The effect of the power, $a, \gamma_{d,\delta}, \gamma_{f,\delta}$, is such that it shifts a little bit $\bar{T}(c)$, yet the behavior of collapsing is seen in the range of powers checked. Increasing the power $a$ increases the influence of the temperature on the stability of the deciders, yet then, the system is more exposed for large stochastic fluctuations. Decreasing $\gamma_{d,\delta}$ helps balance large fluctuations in the temperature, since death happens more often in the camp that increases the temporarily value of the temperature. Nevertheless, note that those with a lot of influence also help balance rapid fluctuations in the temperature. Deceasing $\gamma_{f,\delta}$ has a similar effect when decreasing $\gamma_{d,\delta}$, yet here, an enhanced effect is observed since the fear term are not balanced with other terms.

### B3. Mathematical solutions for the simple model.

In the simplified version, individuals with opinion $\delta$ have the same amount of influence. The deciders obey a simplified equation of motion:

$$\partial_t X_\delta = -k_\delta(t)X_\delta + k_{-\delta}(t)X_{-\delta} \qquad ; \qquad \delta = +,-. \qquad (B.1)$$

Since the dynamics in this model are among opinions, death and birth terms are omitted since these balance each most of the time. (Namely, the above model is conceptually more stable than the extended version.) Note also that we use only linear terms in $X_\delta$; this is done for simplicity, yet also since the term that is not linear in $X_\delta$ in the extended model reflects the existence of the influence coordinate.

The equation of motion for the temperature is a simple deterministic cosine,

$$T(t) = c \cdot \cos(\lambda t). \qquad (B.2)$$

This form is chosen since oscillations in the temperature are clear in all the results we obtained in the extended model. The switching rate in eq. (B.1) follows,

$$k_\delta = k_{\delta,0} e^{\delta T/\bar{T}}, \qquad (B.3)$$

resembling the rates in extended model. We find that here,

$$X_\delta \sim X_\delta(0) \exp\left\{-\delta \frac{c}{\bar{T}} \cos(\lambda t)\right\}.$$

(The mathematical treatment is presebted in what follows.) Clearly, $\frac{c}{\bar{T}} \approx 1$ is the limit of stability in this model. When $\frac{c}{\bar{T}} \gg 1$ oscillations cause $X_\delta$ reaching very high levels or very small levels and so the deciders are not stable. In fact, when starting with 10,000 individuals, the deciders reach very small population's size at $\frac{c}{\bar{T}} = 3$, indicating on collapse. When $\frac{c}{\bar{T}} \ll 1$, the deciders are scarcely influenced when the temperature changes, indicating that the system is not fair (when relating these results with the actual model).

Here, we solve the simple model:

$$\partial_t X_\delta = -k_\delta(t) X_\delta + k_{-\delta}(t) X_{-\delta} \qquad ; \qquad \delta = +, -, \qquad (B.4)$$

with,

$$k_\delta = k_0 e^{\delta T/\bar{T}}, \qquad (B.5)$$

and ,

$$T(t) = c \cdot \cos(\lambda t). \qquad (B.6)$$

First, we use the relation,

$$X_\delta(t) = e^{-\int_0^t k_\delta(a)da} Z_\delta(t), \qquad (B.7)$$

in eq. (B.4), with the conservation of particles, $X_{total} = X_+(t) + X_-(t)$. This results in,

$$\partial_t Z_\delta = e^{\int_0^t k_\delta(a)da} k_{-\delta}(t)(X_{total} - e^{-\int_0^t k_\delta(a)da} Z_\delta).$$

This is rewritten with,

$$\partial_t Z_\delta + k_{-\delta}(t)Z_\delta = e^{\int_0^t k_\delta(a)da}k_{-\delta}(t)X_{total}.$$

writing,

$$Z_\delta(t) = e^{-\int_0^t k_{-\delta}(a)da}Y_\delta(t),$$

we find,

$$\partial_t Y_\delta = e^{\int_0^t k_{total}(q)dq}k_{-\delta}(t)X_{total} \qquad ; \qquad k_{total}(t) = k_+(t) + k_-(t).$$

The solution is,

$$Y_\delta(t) = Y_\delta(0) + X_{total}\int_0^t e^{\int_0^c k_{total}(q)dq}k_{-\delta}(c)dc, \tag{B.8}$$

and so,

$$X_\delta(t) = e^{-\int_0^t k_{total}(q)dq}\left(X_\delta(0) + X_{total}\int_0^t e^{\int_0^c k_{total}(q)dq}k_{-\delta}(c)dc\right). \tag{B.9}$$

Now,

$$\exp\left\{-k_0\int_0^t\left(e^{\left(\frac{c}{T}\right)\cos(\lambda q')} + e^{-\left(\frac{c}{T}\right)\cos(\lambda q')}\right)dq\right\} \sim \exp\left\{-k_0\frac{t}{\pi}(2 + e^{\frac{c}{T}} + e^{-\frac{c}{T}})\right\}.$$

and so,

$$X_\delta(t) \sim X_{total}k_{-\delta}(t) \approx x_{total}\frac{\pi}{2 + e^{\frac{c}{T}} + e^{-\frac{c}{T}}}e^{-\delta\frac{c}{T}\cos(\lambda t)}. \tag{B.10}$$

In this simple model, it is clearly seen that the relative noise $\frac{c}{T}$ determines the stability: when increasing the relative noise, larger fluctuations are seen. When, $\frac{c}{T} \approx 1.55$, the minimal population in a camp is within 10% of the initial value, where when $\frac{c}{T} \approx 3$, we see that the minimal population is within fractions of a percent of the initial value.

**Appendix C: computing the basic relative noise, *fairness* & specific examples**

***An ideal nation.-*** we compute $c_1$ and $\tilde{T}_1$ in an ideal nation. The same computations are used on any nation when computing the basic noise and basic temperature of stability. Note that the expressions for $c_1$ and $\tilde{T}_1$ contain the most stable indicators that characterize all nations and change slowly with time. In the computations of $c$ and $\tilde{T}$ we include the temporary characterization of the nation.

***The basic noise coefficient $c_1$.-*** The basic noise coefficient follows the equation,

$$c_1 = e^{u_{freedom}(u_{individual}+u_{nation}+u_{global})}. \qquad (C.1)$$

Here, averages are not used. We use the actual rate that characterizes the nation. Also, all the

contributions are multiplied with a number $u_{freedom}$. $u_{freedom}$ reflects the freedom in the

nation based on the measure of Freedom House [1]. The index of freedom from the Freedom

House is composed from a rank in the range one and seven in two categories, political rights

and civil liberties, where smaller values reflect a freer nation. We define $u_{freedom}$ with:

$$u_{freedom} = 1 - (-2 + political\ rights + civil\ liberties)/18. \qquad (C.2)$$

Thus, $u_{freedom}$ ranges from 1/3 (for a not free society) and 1 (for a free nation). The other

quantities in eq. (C.1) follows:

(1) $u_{individual}$ is composed from the following indices:

- $u_{unemployment}$ is the monthly unemployment rate and it is computed based on official

figures from the government. In an ideal nation, $u_{unemployment} \leq 0.05$.

- $u_{poverty\ rate}$ is the poverty rate, released from official sources; for example, the US

census bureau in the USA [2]. In an ideal nation, $u_{poverty\ rate} \leq 0.1$

- $1 - HDI$, where $HDI$ is human development index used in the reports of the United

Nations Development Programme [3-4], and it is a number in the interval zero and one, where larger values pointing on a nation with better human development. In an ideal nation, $u_{HDI} = 1 - HDI \leq 0.1$.

- $u_{Gini\ index}$ is the Gini index [5 - 7]. It measures the asymmetry of the wealth distribution in the nation and exists in the unit interval where a larger value indicates on a biased society, relatively. In an ideal nation, $u_{Gini\ index} \leq 0.25$.

- $u_{financial\ markets}$ is the volatility index (VIX) [8]. In an ideal situation, $u_{financial\ markets} \leq 0.1$,

- $u_{protests}$ measures the number of individuals that participate in protests in a year relative with the population's size, where a function that measures the violence of the protest can multiply the number of individuals in any protest in the calculations of $u_{protests}$, where this function ranges from one (a peaceful protest) and 100 (authorities fired on the protesters). In an ideal nation, $u_{protests} \leq 0.01$.

- $u_{opinion\ polls}$ is used when available, where this measures the feeling of the people about their economic conditions, where the value indicates the fraction feeling that they are in poor conditions. See, for example, the statistics publish in Gallup about the situation in the USA [9]. In an ideal nation, $u_{opinion\ polls} \leq 0.1$

- $u_{individual}$ includes all of these contributions with an equal strength, unless additional information suggests that a particular event is the most powerful, for example in times of revolution, e.g., in nations affected from the Arab spring. Thus, basically we write,

$$u_{individual} = \frac{1}{7}(u_{opinion\ polls} + u_{protests} + u_{financial\ markets} + u_{Gini\ index}$$
$$+ u_{HDI} + u_{poverty\ rate} + u_{unemployment}) \quad \text{(C.3.1)}$$

In an ideal nation, $u_{individual} \approx 0.1014$. Note we can use part of the indicators for obtaining $u_{individual}$; for example,

$$u_{individual} = \frac{1}{2}(u_{Gini\ index} + u_{HDI}) \quad \text{(C.3.2)}$$

Equation (C.3.2) is used in computing $c_1$ where other indicators can be used in computing $c$.

- Clearly, when we see mass protests in a nation, all indicators' strength vanish, and we focus on the characterization of the protests, since these are the most accurate message from the citizens of the nation. In fact, $u_{protests}$ is the most significant indicator from all indicators characterizing society, in times of massive protests.

- Other indicators about the wealth in the society can also be used, for example, in times of high inflation or deflation, such indicators are applicable.

(2) $u_{nation}$ is composed from the democracy rate, the debt rate, and the GDP rate.

- $u_{democracy}$ is based on the $democracy\ index$ from the Economist Group [10]. This index aiming at measuring democracy in a nation ranges from zero and 10, where ten is the best rank. The rate is defined with: $u_{democracy} = 1 - democracy\ index/10$. In an ideal nation, $u_{democracy} \leq 0.1$.

- $u_{debt}$ is the governmental debt rate. We use figures from Ref. 6. It is a number larger than zero measuring the debt of the nation in terms of its $GDP$. In an ideal nation, $u_{debt} \leq 0.2$.

- $u_{GDP}$ is the $GDP$ rate. We use figures from ref. 6. It is a number (usually) larger than zero measuring the change in the $GDP$ of the nation, in a year. We define $u_{GDP}$ with, $u_{GDP} = 1 - \%GDP/10$, $\%GDP$ is the percentage of change in the nation's $GDP$. In an ideal nation, $u_{GDP} \leq 0.25$.

- The unbiased $u_{nation}$ thus follows:

$$u_{nation} = \frac{1}{3}(u_{democracy} + u_{debt} + u_{GDP}).$$ (C.4)

In an ideal nation, $u_{nation} \approx 0.18333$.

(3) $u_{Global}$ is composed from the nation's economical ties, its diplomatic ties and conflicts.

- $u_{economical\ ties}$ measures the trading surplus relative with the GDP of the country. We can use data from the WTO website [11]. We define:

$$u_{economical\ ties} = -\frac{export-import}{real\ GDP}.$$
(C.5)

$u_{economical\ ties} \leq 0$ is desirable. Note that among all terms in $u_{Global}$, and also in the other two major categories, this event is the only one that is usually negative. We design this event in such a way so it could balance $u_{con.}$.

- $u_{con.}$ measures the conflicts and wars the nation is involved with:

$$u_{con.} = \mu_{con.,1} conflicts + \mu_{con.,2} distant\ wars\ +\ \mu_{con.,3}\ local\ wars.$$

We suggest: $\mu_{con.,1} = 0.1$, $\mu_{con.,2} = [0.1,1]$ (namely, $0.1 \leq \mu_{con.,2} \leq 1$ depending on the type of the war), $\mu_{con.,3} = [1,10]$, depending on the strength of the war. The value of the event represents the number of such occurrences. Note that $conflicts$ includes territorial disputes and any other war-prone-dispute. In an ideal nation, $u_{conflicts} \leq 0.1$.

- $u_{diplomatic\ ties}$ is the number of nations that are members in the UN that the nation studied has relations with: $u_{diplomatic\ ties} = 1 - diplomatic\ relations/194$. In an ideal nation, $u_{diplomatic\ ties} \leq 0.1$. The idea is that the diplomatic ties increases stability in the entire society and so this term is included here.

- The unbiased $u_{Global}$ thus follows:

$$u_{Global} = \tfrac{1}{3}(u_{economical\ ties} + u_{con.} + u_{diplomatic\ ties}\ ).$$
(C.6)

In an ideal nation, $u_{Global} \approx 0.06$.

- Since $u_{diplomatic\ ties}$ is small for most nations, we may not include it and update the coefficient in (C.6).

- Note that $u_{Global}$ is usually small indicating that the world's nations support stability in the nation.

- Thus, the basic noise in an ideal nation follows, $c_1 = e^{0.3447} = 1.41$

**The basic temperature of stability $\tilde{T}_1$.-** The basic temperature of stability follows the equation,

$$\tilde{T}_1 = e^{q_{freedom}(q_{individual}+q_{nation}+q_{global})}.$$
(C.7)

Here, $q_{freedom}$ is defined with,

$$q_{freedom} = 1 + \frac{1}{6}(-2 + political\ rights + civil\ liberties),$$

where political rights and civil liberties are the measure of the Freedom House [1]. $q_{freedom}$ is 

thus a number ranging from 1 and 3. Most of the events that are part of the quantities in eq.

(C.7) are harder to compute since the idea of a group of deciders controlling the society was not explicitly used in the past, and statistics are scarce. Nevertheless, there are known indicators that point on biases in the society, and we use several of these in the quantification of the categories in eq. (C.7). We suggest the following characterization.

(1) $q_{individual}$ is composed from the following indicators: $q_{wealth}, q_{cor.}, q_{tradition}, q_{pressure}.$

- $q_{wealth}$ measures the total wealth of the top one percent relative with the total wealth

in the nation. We write,

$$q_{wealth} = \frac{\%total\ wealth\ of\ top\ 1pecent}{15} - 1. \qquad (C.8)$$

Statistics about the total wealth of the top one percent in a nation are available [12-13]. Where

these are not known, one can use the statistics about the top 10% [7] and interpolate. Note

that $q_{wealth}$ is (most possibility) in the range, $-1/3 < q_{wealth} \leq 2$. Equation (C.8) is based on

the idea that in a fair society the wealth distribution follows:

6.5%, 8.5%, 18.33%, 13.33%, 53.33% for 0.1%, next 0.9%, next 9%, next 10%, next80%.

The current situation in the USA follows [12]:

17.5%, 17.5%, 35%, 15%,15% for 0.1%, next 0.9%, next 9%, next 10%, next80%.

Basically, the above redistribution of the wealth shifts 21% of the wealth of the top one percent, and 20% of the wealth from the next 19% in favor of the poorer 80% of the people,

relative with the current situation in the USA. The top one percent loses about 60% from its

power, and the next 19% loses about 40% from its power, yet, wins power against the top one

percent, relative with the current situation in the USA. Basically, in the ideal nation, the poorer

80% has about 53% of the wealth and thus has control, at least in terms of wealth. We further talk about this point in the **General conclusions** part in this appendix. Here we note that there are several options for redistribution of the wealth. In all of the, the top one percent loses wealth, and other 99% win wealth.

- $q_{tradition}$ measures the bias the deciders have due to tradition in the nation. It is affected from several issues: (1) historical considerations, e.g. world wars, long lasting conflicts, long lasting common wisdom, etc. (2) the existence of a royal society and other hierarchical societies that affect the nation, e.g. military, religion, etc. (3) The level of blind admiration for fame and money in the nation. In a fair society, $q_{tradition} \approx 2/3$.

- Other measures that in principle can be used although are harder to calculate are the following: (1) $q_{cor.}$ measures the corruption in the nation. We can look on the number of high rank individuals that are reportedly related with corruption per year, relative with an expected number of such cases. The difficulty here is that in a democratic nation, such reports are much more abundant than in a nation that is not democratic, although the actual level of corruption is perhaps much higher in nations that are not democratic. (2) $q_{opportunities}$ reflects the opportunities or the flow inside the top one in a thousand. For example, one can look on the flow of individuals in high rank positions in the media. A simple analysis may reveal the lack of such opportunities. (3) $q_{pressure}$ measures the pressure put on the deciders from the media and the authorities.

- When specific information on the indicators presented in the last point is absent, we write,

$$q_{individual} = \frac{1}{2}(q_{wealth} + q_{tradition}), \tag{C.9}$$

where we can include other the indicators in a manner of an unbiased average and a biased average. Basically, eq. (C.9) is used in the computation of $\tilde{T}_1$, where other terms can appear in the computation of $\tilde{T}$. Note that $q_{tradition}$ is at least one in most cases and $q_{individual}$ is in many cases larger than one.

- In an ideal nation, $q_{individual} = \frac{1}{2}(0 + 2/3) = 1/3$.

(2) $q_{nation}$ includes all the regulations and laws that directly affect the deciders.

- $q_{laws}$ measure considers indicators about the nation's laws that affect the deciders. We write:

$$q_{laws} = \frac{4}{3} u_{democracy}. \tag{C.10}$$

Note that $q_{laws}$ is in the range, $0 \leq q_{laws} \leq 4/3$. The main reason that the number $\frac{4}{3}$ appears in (C.16) is that the total strength of $u_{democracy}$ in the relative noise is one in a free society, since $u_{democracy}$ appears also in (C.4).

- We can include other indicators that reflect the strength of the regulations in the nation with regard of favoring the deciders. These are included in the computation of $\tilde{T}$. In particular, we suggest using $q_{new\ regulations}$ that measures the strength and importance of new regulations that affect the deciders' strength.

- Basically we write,

$$q_{nation} = q_{laws}. \tag{C.11}$$

- In an ideal nation, $q_{nation} \leq 0.133$.

(3) $q_{global}$ includes any contribution from the world that affects the deciders.

- $q_{strength}$ is defined with the measure in ref. 14, where countries are either Supper powers, Great powers, Regional powers, Middle powers, or, we define, other independent nations with little power. We rank these, 5, 4, 3, 2, 1, respectively. This indicator measures political, cultural and economic influence of a nation on other nations. We write,

$$q_{strength} = rank/5. \tag{C.12}$$

Note that $q_{strength}$ ranges, $0 < q_{strength} \leq 1$.

- Other measures that may contribute are the following: (1) $q_{IMF}$ is the amount of shares the nation has in the IMF [15] and measures the financial power the nation has among the world's nations and thus may help supporting the deciders in internal matters also, similar with the indicator about the strength of the nation. We can write,

$$q_{IMF} = n_{IMF}/\max_n(n_{IMF}), \qquad\qquad (C.13)$$

where $n_{IMF}$ is the number of shares the nation has in the IMF, and we look on the ratio relative with the maximal shares a country has in the IMF (USA with about 16% of the shares.) Equation (C.13) measures economical influence in the world. It is more accurate than a five-step scale since any nation has shares in the IMF and it is a continuous number in principle, so the scale is more sensitive. Note that $q_{IMF}$ ranges, $0 \leq q_{IMF} \leq 1$. (2) $q_{specific\ ties}$, when known, measures the amount of economical ties the deciders have with other nations. (3) $q_{sanctions}$ measures the amount of sanctions applied on the deciders from the world's nations

- When other indicators are not known, $q_{global}$ follows,

$$q_{global} = \frac{1}{2}(q_{IMF} + q_{strength}). \qquad\qquad (C.14)$$

Basically, eq. (C14) represents $q_{global}$ in $\widetilde{T}_1$, where other indicators can appear in $\tilde{T}$.

- In an ideal nation, $q_{global} \leq 0.5$. The internal contribution of the various terms can vary.

- In an ideal nation,

$$\widetilde{T}_1 \leq e^{1*(0.5+0.133+1/3)} = 2.63. \qquad\qquad (C.15)$$

- Thus, the relative noise in an ideal nation follows,

$$\frac{c_1}{\widetilde{T}_1} = 0.536. \qquad\qquad (C.16)$$

**The fairness.-** The basic fairness in a nation is defined with,

$$fairness = \frac{\widetilde{T}_{1,ideal}}{\widetilde{T}_1}. \qquad\qquad (C.17)$$

$\tilde{T}_{1,ideal}$ is defined in equation (C.15). $fairness$ is a number in the range, zero and one, where a larger value indicates on a nation that is relative more fair. The temporarily fairness uses $\tilde{T}$ instead of $\tilde{T}_1$:

$$temporarily\ fairness = \frac{\tilde{T}_{1,ideal}}{\tilde{T}}. \qquad (C.18)$$

***Specific examples: computations of $c_1$, $\tilde{T}_1$, fairness and $c/\tilde{T}$:***

***A general note***: We use the CIA fact book [6] and find the income for the top 10% in a nation. We then use the figures about the USA and obtain the wealth of the top 1%. In the USA, 30% of the income is for top 10% (6). Yet, in ref. 12, the total wealth of the top 1 percent in the USA is 34.6%. Thus, the income figure is multiply with 1.1533, and we obtain the top 1% in a nation.

- ***China: $c_1 \approx 1.1948, \tilde{T}_1 \approx 241.16$, $fairness = 0.01$***

$\tilde{T}_1 \approx e^{2.83*1.938} = 241.16$: (1) $q_{freedom} = 1 + \frac{1}{6}(-2 + political\ rights + civil\ liberties) = 2.83$. (2) $q_{wealth} \approx \frac{15}{15} - 1 = 0$. (3) $q_{tradition} \approx 1$. Since China tradition biases in favor of those in power. (4) $q_{individual} \approx \frac{1}{2}(0 + 1) = 1/2$. (5) $q_{nation} \approx \frac{4}{3}0.686 = 0.91467$. (6) $q_{global} \approx \frac{1}{2}\left(\frac{3.81}{16.76} + \frac{4}{5}\right) = 0.52366$

$c_1 \approx e^{0.388*0.4577} = 1.1948$: (1) $u_{freedom} = 1 - \frac{-2 + political\ rights + civil\ liberties}{18} = 0.388$, (2) $u_{individual} \approx \frac{1}{2}(u_{Gini\ index} + u_{HDI}) = \frac{1}{2}(0.415 + 0.337) = 0.376$. (3) $u_{nation} = \frac{1}{3}(u_{democracy} + u_{debt} + u_{GDP}) = \frac{1}{3}(0.686 + 0.189 - 0.03) = 0.28167$. (3) $u_{Global} \approx u_{economical\ ties} = -0.2$.

***Current situation:*** Protests emerged recently in China (the Jasmine revolution, 2011), and were oppressed. In terms of the model both the noise and the temperature of stability increased. China should shift in the direction of a free and democratic society and reduce $\tilde{T}$.

- ***Russia: $c_1 \approx 2.1349, \tilde{T}_1 \approx 406.43, fairness = 0.0065$***

$\widetilde{T}_1 \approx e^{2.5*2.4} = \mathbf{406.43}$: (1) $q_{freedom} = 1 + \frac{1}{6}(-2 + political\ rights + civil\ liberties) = 2.5$. (2) $q_{wealth} \approx \frac{35}{15} - 1 = \frac{4}{3}$. (3) $q_{tradition} \approx 1$. Since Russia tradition biases in favor of those in power. (4) $q_{individual} \approx \frac{1}{2}\left(\frac{4}{3} + 1\right) = 7/6$. (5) $q_{nation} \approx \frac{4}{3}0.574 = 0.765$. (6) $q_{global} \approx \frac{1}{2}\left(\frac{2.39}{16.76} + \frac{4}{5}\right) = 0.4713$.

$c_1 \approx e^{0.5*1.5168} = \mathbf{2.1349}$: (1) $u_{freedom} = 1 - \frac{-2 + political\ rights + civil\ liberties}{18} = 0.5$. (2) $u_{individual} \approx \frac{1}{2}(u_{Gini\ index} + u_{HDI}) = 0.5(0.45 + 0.281) = 0.3655$. (3) $u_{nation} = \frac{1}{3}\left(u_{democracy} + u_{debt} + u_{GDP}\right) = \frac{1}{3}(0.574 + 0.09 + 0.6) = 0.4213$. (4) $u_{Global} \approx u_{economical\ ties} = 0.73$.

***Current situation:*** Protests emerged recently in Russia and were oppressed (2011). In terms of the model both the noise and the temperature of stability increased. Russia should shift in the direction of a free and democratic society and reduce $\widetilde{T}$.

- ***Japan:*** $c_1 \approx \mathbf{2.69}, \widetilde{T}_1 \approx \mathbf{8.58}, fairness = \mathbf{0.306}$

$\widetilde{T}_1 \approx e^{1.167*1.842} = \mathbf{8.58}$: (1) $q_{freedom} = 1 + \frac{1}{6}(-2 + political\ rights + civil\ liberties) = 1.167$. (2) $q_{wealth} \approx \frac{30}{15} - 1 = 1$. (3) $q_{tradition} \approx 1$. Japan has an emperor, and is a very hierarchical society. Due to WWII, the value of $q_{tradition}$ chosen here is one, rather than a higher value. (4) $q_{individual} \approx \frac{1}{2}(1 + 1) = 1$. (5) $q_{nation} \approx \frac{4}{3}0.192 = 0.256$. (6) $q_{global} \approx \frac{1}{2}\left(\frac{6.24}{16.76} + \frac{4}{5}\right) = 0.586$.

$c_1 \approx e^{0.9444*1.049} = \mathbf{2.69}$: (1) $u_{freedom} = 1 - \frac{-2 + political\ rights + civil\ liberties}{18} = 0.9444$. (2) $u_{individual} \approx \frac{1}{2}(u_{Gini\ index} + u_{HDI}) = 0.5(0.376 + 0.116) = 0.246$. (3) $u_{nation} = \frac{1}{3}\left(u_{democracy} + u_{debt} + u_{GDP}\right) = \frac{1}{3}(0.192 + 1.975 + 0.61) = 0.9257$. (4) $u_{Global} \approx u_{economical\ ties} = -0.1229$.

***Current situation:*** The noise term should contain at least an additional *e* resulting from the Tsunami and the nuclear crisis. Although the society does not immediately blame the deciders in cases of natural disasters, the government is currently unstable in Japan, contributing to the public anger. In term of the society-deciders model, the relative noise in Japan is currently about one, $\frac{c}{T} \approx 1$. Thus, the people can demand serious economical amendments, shifting wealth from the top one percent in favor of the poorest 80 percent.

- **Sweden:** $c_1 \approx 1.52, \widetilde{T}_1 \approx 3.13, fairness = 0.84$

$\widetilde{T}_1 \approx e^{1.14} = 3.13$: (1) $q_{freedom} = 1$. (2) $q_{wealth} \approx \frac{25}{15} - 1 = \frac{2}{3}$. (3) $q_{tradition} \approx 1$. Since Sweden has a royal society. (4) $q_{individual} \approx \frac{1}{2}\left(\frac{2}{3} + 1\right) = 0.833$. (5) $q_{nation} \approx \frac{4}{3} 0.05 = 0.067$. (6) $q_{global} \approx \frac{1}{2}\left(\frac{1.4}{16.76} + \frac{2}{5}\right) = 0.2418$

$c_1 \approx e^{0.40} = 1.52$: (1) $u_{freedom} = 1$. (2) $u_{individual} \approx \frac{1}{2}(u_{Gini\ index} + u_{HDI}) = 0.5(0.23 + 0.115) = 0.1725$. (3) $u_{nation} = \frac{1}{3}(u_{democracy} + u_{debt} + u_{GDP}) = \frac{1}{3}(0.05 + 0.39 + 0.45) = 0.299$. (4) $u_{Global} \approx u_{economical\ ties} = -0.0517$.

*Current situation:* The problem of the wealth distribution appears also in Sweden where the poorer 80% hold about 25% of the wealth. Within the model, we find that Sweden basic relative noise follows, $c_1/\widetilde{T}_1 \approx 0.4856$. Thus Sweden should try reducing $\widetilde{T}$ while shifting wealth in the direction of the poorer 80% with new laws and regulations.

- **Germany:** $c_1 \approx 1.96, \widetilde{T}_1 \approx 3.92, fairness = 0.67$

$\widetilde{T}_1 \approx e^{1.366} = 3.92$: (1) $q_{freedom} = 1$. (2) $q_{wealth} \approx \frac{25}{15} - 1 = \frac{2}{3}$. (3) $q_{tradition} \approx 2/3$. Since Germany was an empire through its history. WWII decreases the value of $q_{tradition}$ in about one third. (4) $q_{individual} \approx \frac{1}{2}\left(\frac{2}{3} + \frac{2}{3}\right) = 2/3$. (5) $q_{nation} \approx \frac{4}{3} 0.162 = 0.216$. (6) $q_{global} \approx \frac{1}{2}\left(\frac{5.81}{16.76} + \frac{4}{5}\right) = 0.4838$

$c_1 \approx e^{0.6735} = 1.96$: (1) $u_{freedom} = 1$. (2) $u_{individual} \approx \frac{1}{2}(u_{Gini\ index} + u_{HDI}) = 0.5(0.27 + 0.115) = 0.1925$. (3) $u_{nation} = \frac{1}{3}(u_{democracy} + u_{debt} + u_{GDP}) = \frac{1}{3}(0.162 + 0.832 + 0.65) = 0.548$. (4) $u_{Global} \approx u_{economical\ ties} = -0.067$.

*Current situation:* The problem of the wealth distribution appears also in Germany, where the poorer 80% holds about 35% of the wealth. Although this is considered a better situation relative with our western nations, Germany basic relative noise is still small, $c_1/\tilde{T}_1 \approx 0.5$. In addition, there are local economical woes resulting from the EU debt crisis (2010-2011). Thus, Germany should try reducing $\tilde{T}$ while shifting wealth in the direction of the poorer 80% with new laws and regulations, and increasing the dynamics among socio economic classes, specifically inside the top five percent and the top one percent.

- **_France: $c_1 \approx 2.75$, $\tilde{T}_1 \approx 6$, $fairness = 0.438$_**

$\tilde{T}_1 \approx e^{1.79} = 8.77$: (1) $q_{freedom} = 1$. (2) $q_{wealth} \approx \frac{29}{15} - 1 = 0.933$. (3) $q_{tradition} \approx 1$. Since France was an empire through its history. (4) $q_{individual} \approx \frac{1}{2}(0.933 + 1) = 0.967$. (4) $q_{nation} \approx \frac{4}{3}0.223 = 0.2973$. (5) $q_{global} \approx \frac{1}{2}\left(\frac{4.29}{16.76} + \frac{4}{5}\right) = 0.52798$

$c_1 \approx e^{1.0129} = 2.7537$: (1) $u_{freedom} = 1$. (2) $u_{individual} \approx \frac{1}{2}(u_{Gini\ index} + u_{HDI}) = 0.5(0.327 + 0.128) = 0.2275$. (3) $u_{nation} = \frac{1}{3}(u_{democracy} + u_{debt} + u_{GDP}) = \frac{1}{3}(0.223 + 0.824 + 0.85) = 0.6323$. (4) $u_{Global} \approx u_{economical\ ties} = 0.1531$.

**_Current situation:_** The problem of the wealth distribution appears also in France where the poorer 80% hold only about 25% of the wealth. Within the model, we find that France basic relative noise follows, $c_1/\tilde{T}_1 \approx 0.458$. We also heard about new economical regulations (2010-2011). Still, France should try reducing $\tilde{T}$ while shifting wealth in the direction of the poorer 80% with new laws and regulations.

- **_UK: $c_1 \approx 2.4244$, $\tilde{T}_1 \approx 5.89$, $fairness = 0.447$_**

$\tilde{T}_1 \approx e^{1.77} = 5.89$: (1) $q_{freedom} = 1$. (2) $q_{wealth} \approx \frac{30}{15} - 1 = 1$. (3) $q_{tradition} \approx 1$. Since UK was an empire through its history and has a royal society. (4) $q_{individual} \approx \frac{1}{2}(1 + 1) = 1$. (5) $q_{nation} \approx \frac{4}{3}0.184 = 0.24533$. (6) $q_{global} \approx \frac{1}{2}\left(\frac{4.29}{16.76} + \frac{4}{5}\right) = 0.527$

$c_1 \approx e^{0.8856} = 2.4244$: (1) $u_{freedom} = 1$. (2) $u_{individual} \approx \frac{1}{2}(u_{Gini\ index} + u_{HDI}) = 0.5(0.34 + 0.151) = 0.2455$. (3) $u_{nation} = \frac{1}{3}(u_{democracy} + u_{debt} + u_{GDP}) = \frac{1}{3}(0.184 + 0.761 + 0.87) = 0.605$. (4) $u_{Global} \approx u_{economical\ ties} = 0.0351$.

***Current situation:*** The problem of the wealth distribution appears also in the UK where the poorer 80% hold only about 35% of the wealth. We see protests recently in the UK (2011). We also heard about the austerity measures (2010-2011) due to the economical stress. Within the model, we find that UK basic relative noise follows, $c_1/\widetilde{T}_1 \approx \mathbf{0.41159.}$ The UK should try reducing $\widetilde{T}$ while shifting wealth in the direction of the poorer 80% with new laws and regulations; namely, the austerity measure should help in this direction rather than in the opposite direction.

- ***Belgium:*** $c_1 \approx \mathbf{2.3985}$ **,** $\widetilde{T}_1 \approx \mathbf{4.55,}$ $\boldsymbol{fairness} = \mathbf{0.58}$

$\widetilde{T}_1 \approx e^{\mathbf{1.515}} = \mathbf{5.694}$: (1) $q_{freedom} = 1$. (2) $q_{wealth} \approx \frac{30}{15} - 1 = 1$. (3) $q_{tradition} \approx 1$. Since Belgium has a royal society. (4) $q_{individual} \approx \frac{1}{2}(1+1) = 1$. (5) $q_{nation} \approx \frac{4}{3}0.195 = 0.26$. (6) $q_{global} \approx \frac{1}{2}\left(\frac{1.86}{16.76} + \frac{2}{5}\right) = 0.255$

$c_1 \approx e^{\mathbf{0.8748}} = \mathbf{2.3985}$: (1) $u_{freedom} = 1$. (2) $u_{individual} \approx \frac{1}{2}(u_{Gini\ index} + u_{HDI}) = 0.5(0.28 + 0.133) = 0.2065$. (3) $u_{nation} = \frac{1}{3}(u_{democracy} + u_{debt} + u_{GDP}) = \frac{1}{3}(0.195 + 1.01 + 0.8) = 0.6683$. (4) $u_{Global} \approx u_{economical\ ties} = 0$.

***Current situation:*** The problem of the wealth distribution appears also in the Belgium where the poorer 80% hold only about 35% of the wealth. Within the model, we find that Belgium basic relative noise follows, $\frac{c_1}{\widetilde{T}_1} \approx \mathbf{0.527.}$ During 2011, Belgium political system was in stagnation, where parties could not form a government. Belgium has a problem with its population (Dutch speaking and France speaking individuals), where one part wants establishing a new nation. Within the model, and with a united Belgium, Belgium should try reducing $\widetilde{T}$ while shifting wealth in the direction of the poorer 80% with new laws and regulations, and solve its internal differences.

- ***Spain:*** $c_1 \approx \mathbf{2.2427}$ **,** $\widetilde{T}_1 \approx \mathbf{4.54,}$ $\boldsymbol{fairness} = \mathbf{0.58}$

$\widetilde{T}_1 \approx e^{1.512} = 4.5386$: (1) $q_{freedom} = 1 + \frac{1}{6}(-2 + political\ rights + civil\ liberties) = 1.$

(2) $q_{wealth} \approx \frac{27.57}{15} - 1 = 0.838.$ (3) $q_{tradition} \approx 1.$ Spain has a royal society. (4) $q_{individual} \approx \frac{1}{2}(0.838 + 1) = 0.919.$ (5) $q_{nation} \approx \frac{4}{3}0.184 = 0.245.$ (6) $q_{global} \approx \frac{1}{2}\left(\frac{1.63}{16.76} + \frac{3}{5}\right) = 0.348$

$c_1 \approx e^{0.8077} = 2.2427$: (1) $u_{freedom} = 1 - \frac{-2 + political\ rights + civil\ liberties}{18} = 1.$ (2) $u_{individual} \approx \frac{1}{2}(u_{Gini\ index} + u_{HDI}) = 0.5(0.32 + 0.137) = 0.2285.$ (3) $u_{nation} = \frac{1}{3}(u_{democracy} + u_{debt} + u_{GDP}) = \frac{1}{3}(0.184 + 0.6 + 1.1) = 0.628.$ (4)

$u_{Global} \approx u_{economical\ ties} = 0.048.$

**Current situation:** The noise term should contain at least an additional *e* resulting from the high unemployment rate and protests (2011). This means that the relative noise is about one, $\frac{c}{T} \approx 1.$ This is an interesting situation where the people can demand serious economical amendments from the deciders, practically shifting wealth from the top one percent in favor of the poorest 80 percent.

- **Greece:** $c_1 \approx 3.04, \widetilde{T}_1 \approx 5.16, fairness = 0.51$

$\widetilde{T}_1 \approx e^{\frac{7}{6}*1.406} = 5.16$: (1) $q_{freedom} = 1 + \frac{1}{6}(-2 + political\ rights + civil\ liberties) = 7/6.$

(2) $q_{wealth} \approx \frac{27.58}{15} - 1 = 0.83867.$ (3) $q_{tradition} \approx 1.$ Greece was an ancient empire. (4) $q_{individual} \approx \frac{1}{2}(1 + 1) = 0.919.$ (5) $q_{nation} \approx \frac{4}{3}0.208 = 0.2733.$ (6) $q_{global} \approx \frac{1}{2}\left(\frac{0.47}{16.76} + \frac{2}{5}\right) = 0.214$

$c_1 \approx e^{0.9444*1.29617} = 3.037$: (1) $u_{freedom} = 1 - \frac{-2 + political\ rights + civil\ liberties}{18} = 0.9444.$ (2) $u_{individual} \approx \frac{1}{2}(u_{Gini\ index} + u_{HDI}) = 0.5(0.33 + 0.145) = 0.2375.$ (3) $u_{nation} = \frac{1}{3}(u_{democracy} + u_{debt} + u_{GDP}) = \frac{1}{3}(0.208 + 1.428 + 1.45) = 1.02867.$ (4) $u_{Global} \approx u_{economical\ ties} = 0.03.$

**Current situation:** The noise term should contain at least an additional *e* resulting from the protests and the increasing debt problem (2011). This means that the relative noise is about one, $\frac{c}{T} \approx 1.$ This is an interesting situation where the people can demand serious economical amendments from the deciders, practically shifting wealth from the top one percent in favor of the poorest 80 percent.

- **Egypt: $c_1 \approx 1.465, \tilde{T}_1 \approx 17.1, fairness = 0.154$**

$\tilde{T}_1 \approx e^{1.833*1.5486} = 17.1$: (1) $q_{freedom} = 1 + \frac{1}{6}(-2 + political\ rights + civil\ liberties) = 1.833$. Since Egypt is in the final stages of a revolution moving in the direction for democracy, we use: $political\ rights = 3$; $civil\ liberties = 3$. (2) $q_{wealth} \approx \frac{29.3}{15} - 1 = 1$. (3) $q_{tradition} \approx 1$. Pharaohs' tradition. (4) $q_{individual} \approx \frac{1}{2}(0.9533 + 1) = 0.9767$. (5) $q_{nation} \approx \frac{4}{3}0.27 = 0.36$

Here we used, $u_{democracy} = 0.27$, resulting from the revolution. This value points on, $democracy = 7.3$, in the index of democracy in Ref. 11, and is about the top of the range of a democracy of the second rank. (6) $q_{global} \approx \frac{1}{2}\left(\frac{0.40}{16.76} + \frac{2}{5}\right) = 0.2119$

$c_1 \approx e^{0.777*0.49} = 1.465$: (1) $u_{freedom} = 1 - \frac{-2 + political\ rights + civil\ liberties}{18} = 0.777$. (2) $u_{individual} \approx \frac{1}{2}(u_{Gini\ index} + u_{HDI}) = 0.5(0.321 + 0.38) = 0.3505$. (3) $u_{nation} = \frac{1}{3}(u_{democracy} + u_{debt} + u_{GDP}) = \frac{1}{3}(0.27 + 0.799 + 0.49) = 0.51967$. (4) $u_{Global} \approx u_{economical\ ties} = -0.379$. (5)

**The current situation:** Protests emerge recently in Egypt (December 2010). These were developed in the direction of a revolution against the Mubarak regime. It seems that the revolution is reaching its initial goals, and democratic elections are scheduled (Winter 2011). Egypt should continue shifting in the direction of a free and democratic society and reduce $\tilde{T}$.

- **Libya: $c_1 \approx 1.94, \tilde{T}_1 \approx 18.54, fairness = 0.142$**

$\tilde{T}_1 \approx e^{1.833*1.59} = 18.54$: (1) $q_{freedom} = 1 + \frac{1}{6}(-2 + political\ rights + civil\ liberties) = 1.833$. Since Libya is in the final stages of a revolution moving in the direction for democracy, we use: $political\ rights = 3$; $civil\ liberties = 3$. (2) $q_{wealth} \approx not\ known$. (3) $q_{tradition} \approx 1$. A tribal society. (4) $q_{individual} \approx 1$. (5) $q_{nation} \approx \frac{4}{3}0.3588 = 0.4784$. Here we used, $u_{democracy} = 0.3588$, resulting from the revolution. This value points on a $democracy =$

6.412 in the democracy index ref. 10, and is about the middle of the range of a democracy of the second rank. (6) $q_{global} \approx \frac{1}{2}\left(\frac{0.48}{16.76} + \frac{1}{5}\right) = 0.1143$

$c_1 \approx e^{0.777*(u_{individual}+u_{nation})*3/2} = \mathbf{1.94}$: (1) $u_{freedom} = 1 - \frac{-2+political\ rights+civil\ liberties}{18} =$

0.777. (2) $u_{individual} \approx u_{HDI} = 0.245$. (3) $u_{nation} = \frac{1}{3}(u_{democracy} + u_{debt} + u_{GDP}) =$
$\frac{1}{3}(0.3588 + 0.033 + 0.58) = 0.324$. (4) $u_{Global} \approx not\ known$.

***The current situation:*** Protests emerge recently in Libya (Spring 2011). These were developed in the direction of a revolution against the Kaddafi regime. It seems that the revolution is reaching its goals (Winter 2011). Libya should continue shifting in the direction of a free and democratic society and reduce $\widetilde{T}$.

- ***Syria:*** $c_1 \approx \mathbf{1.852}, \widetilde{T}_1 \approx \mathbf{601.53}, fairness = \mathbf{0.0044}$

$\widetilde{T}_1 \approx e^{2.833*2.2586} = \mathbf{601.53}$: (1)

$q_{freedom} = 1 + \frac{1}{6}(-2 + political\ rights + civil\ liberties) = 2.833$. (2) $q_{wealth} \approx$
$not\ known$. (3) $q_{tradition} \approx 1$. The minority group, Alawis, controls of the people. (4)
$q_{individual} \approx 1 = 1$. (5) $q_{nation} \approx \frac{4}{3}0.869 = 1.158$. (6) $q_{global} \approx \frac{1}{2}\left(\frac{0.15}{16.76} + \frac{1}{5}\right) = 0.1$

$c_1 \approx e^{0.444*1.388} = \mathbf{1.852}$: (1) $u_{freedom} = 1 - \frac{-2+political\ rights+civil\ liberties}{18} = 0.3888$. (2)
$u_{individual} \approx \frac{1}{2}(u_{Gini\ index} + u_{HDI}) = 0.5(0.4 + 0.411) = 0.4055$. Here, we use a bound for the Gini index, $u_{Gini\ index} \leq 0.4$, based on other nations, e.g., Iran. (3)
$u_{nation} = \frac{1}{3}(u_{democracy} + u_{debt} + u_{GDP}) = \frac{1}{3}(0.869 + 0.286 + 0.68) = 0.61167$. (4)
$u_{Global} \approx \frac{1}{2}(u_{economical\ ties} + u_{con.}) = \frac{1}{2}(0.442 + 0.3) = 0.371$.

***The current situation:*** Protests emerge recently in Syria (Spring 2011). The protests are still going on, but the regime oppresses them. In terms of the model, both the noise and the temperature of stability increased. Sanctions were put on Syria from the EU and USA (Summer-Fall 2011), decreasing $\widetilde{T}$. Syria should shift in the direction of a free and democratic society and reduce $\widetilde{T}$.

- **Saud Arabia: $c_1 \approx 1.527, \tilde{T}_1 \approx 819.96, fairness = 0.0032$**

$\tilde{T}_1 \approx e^{2.833*2.368} = 819.96$: (1)

$q_{freedom} = 1 + \frac{1}{6}(-2 + political\ rights + civil\ liberties) = 2.8333$. (2) $q_{wealth} \approx$

$not\ known$. (3) $q_{tradition} \approx 1$. Saud Arabia has a royal society. (4) $q_{individual} \approx 1$. (5)

$q_{nation} \approx \frac{4}{3} u_{democracy} = 1.088$. (6) $q_{global} \approx \frac{1}{2}\left(\frac{2.8}{16.76} + \frac{2}{5}\right) = 0.28$

$c_1 \approx e^{0.3888*1.089} = 1.527$: (1) $u_{freedom} = 1 - \frac{-2+political\ rights+civil\ liberties}{18} = 0.3888$. (2)

$u_{individual} \approx u_{HDI} = 0.25$. (3) $u_{nation} = \frac{1}{3}(u_{democracy} + u_{debt} + u_{GDP}) = \frac{1}{3}(0.816 + 0.171 +$

$0.63\ ) = 0.539$. (4) $u_{Global} \approx u_{con.} = 0.3$.

**Current situation:** The noise term should contain at least an additional *e* resulting from recent

protests (Spring 2011), and the temperature of stability should contain an additional term of

$1/e$ since new regulations in favor of democracy were declared recently in Saud Arabia

(Summer-Fall 2011). Still, the relative noise in extremely small, $\frac{c}{\tilde{T}} \approx 1/60$. Saud Arabia should decrease $\tilde{T}$ while moving in the direction of democracy.

- **Iran: $c_1 \approx 1.98, \tilde{T}_1 \approx 1329, fairness = 0.00198$**

$\tilde{T}_1 \approx e^{2.67*2.694} = 1329$: (1) $q_{freedom} = 1 + \frac{1}{6}(-2 + political\ rights + civil\ liberties) =$

$2.67$. (2) $q_{wealth} \approx \frac{34}{15} - 1 = 1.27$. (3) $q_{tradition} \approx \frac{4}{3}$. The religious group has a serious influence on the perception of the people. (4) $q_{individual} \approx \frac{1}{2}\left(\frac{4}{3} + 1.27\right) = 1.3$. (5) $q_{nation} \approx \frac{4}{3}0.806 =$

$1.074$. (6) $q_{global} \approx \frac{1}{2}\left(\frac{0.62}{16.76} + \frac{3}{5}\right) = 0.318$

$c_1 \approx e^{0.444*1.5363} = 1.98$: (1) $u_{freedom} = 1 - \frac{-2+political\ rights+civil\ liberties}{18} = 0.444$. (2)

$u_{individual} \approx u_{Gini\ index} = 0.445$. (3) $u_{nation} = \frac{1}{3}(u_{democracy} + u_{debt} + u_{GDP}) = \frac{1}{3}(0.806 +$

$0.166 + 0.9\ ) = 0.624$. (4) $u_{Global} \approx \frac{1}{2}(u_{economical\ ties} + u_{con.}) = \frac{1}{2}(0.6346 + 0.3) = 0.4673$.

**The current situation:** Protests emerge recently in Iran (summer 2009) and were oppressed. In terms of the model both the noise and the temperature of stability increased. Iran should shift in the direction of a free and democratic society and reduce $\tilde{T}$.

- **Israel: $c_1 \approx 2.51, \widetilde{T}_1 \approx 5.92, fairness = 0.3986$**

$\widetilde{T}_1 \approx e^{1.167*1.524} = 5.92$: (1) $q_{freedom} = 1 + \frac{1}{6}(-2 + political\ rights + civil\ liberties) = 1.167$. (2) $q_{wealth} \approx \frac{27}{15} - 1 = 0.8$. (3) $q_{tradition} \approx 1.15$. The military has a serious influence on the perception of the people. Fame and money are over-appreciated. (4) $q_{individual} \approx \frac{1}{2}(1.15 + 0.8) = 0.975$. (5) $q_{nation} \approx \frac{4}{3}0.252 = 0.336$. (6) $q_{global} \approx \frac{1}{2}\left(\frac{0.45}{16.76} + \frac{2}{5}\right) = 0.213$

$c_1 \approx e^{0.9444*0.9723} = 2.51$: (1) $u_{freedom} = 1 - \frac{-2 + political\ rights + civil\ liberties}{18} = 0.9444$. (2) $u_{individual} \approx \frac{1}{2}(u_{Gini\ index} + u_{HDI}) = 0.5(0.392 + 0.128) = 0.26$. (3) $u_{nation} = \frac{1}{3}\left(u_{democracy} + u_{debt} + u_{GDP}\right) = \frac{1}{3}(0.252 + 0.745 + 0.54) = 0.512$. (4) $u_{Global} \approx \frac{1}{2}(u_{economical\ ties} + u_{con.}) = -0.1 + 0.3 = 0.2$.

***The current situation:*** Protests emerge recently in Israel (Summer 2011) due to the distorted wealth distribution. A committee was formed by the government and suggested a series of economical amendments. In terms of the model, the noise increased. The new regulations should put in place for reducing $\widetilde{T}$ and creating a fair society. The problem is that in Israel the condition with the Palestinians can create a war, in which the temperature of stability of the deciders immediately soars, and measures increasing fairness in the society are pushed aside. The recommendation is thus making peace with the Palestinians and design new regulations for decreasing $\widetilde{T}$ while moving in the direction of a more fair society.

- **USA: $c_1 \approx 2.39, \widetilde{T}_1 \approx 12.93, fairness = 0.2$**

$\widetilde{T}_1 \approx e^{2.559} = 12.933$: (1) $q_{freedom} = 1 + \frac{1}{6}(-2 + political\ rights + civil\ liberties) = 1$. (2) $q_{wealth} \approx \frac{34.6}{15} - 1 = 1.31$. (3) $q_{individual} \approx \frac{1}{2}\left(\frac{4}{3} + 1.3\right) = 1.3167$. (4) $q_{nation} \approx \frac{4}{3}0.182 = 0.243$. (5) $q_{global} \approx \frac{1}{2}(1 + 1) = 1$

$c_1 \approx e^{1.01} = 2.39$: (1) $u_{freedom} = 1 - \frac{-2 + political\ rights + civil\ liberties}{18} = 1$, (2) $u_{individual} \approx \frac{1}{2}(u_{Gini\ index} + u_{HDI}) = 0.5(0.45 + 0.098) = 0.274$. (3) $u_{nation} = \frac{1}{3}\left(u_{democracy} + u_{debt} + \right.$

$u_{GDP}) = \frac{1}{3}(0.182 + 0.623 + 0.72) = 0.508.$ (4) $u_{Global} \approx \frac{1}{2}(u_{economical\ ties} + u_{con.}) = \frac{-0.069+0.3}{2} = 0.115.$

***The current situation:*** Protests emerge recently in USA (September 2011) due to the distorted wealth distribution and the high unemployment rate. In terms of the model the noise increased. New regulations should put in place for reducing $\widetilde{T}$ and creating a fair society. The main idea is shifting the wealth from the top one percent and also from the next 19% in the direction of the poorest 80%. Indeed, the political motivation for this is created with public awakening, namely, with mass protests.

- ***Brazil:*** $c_1 \approx 1.989, \widetilde{T}_1 \approx 19.55, fairness = 0.135$

$\widetilde{T}_1 \approx e^{1.33*2.33} = 19.55$: (1) $q_{freedom} = 1 + \frac{1}{6}(-2 + political\ rights + civil\ liberties) = \frac{4}{3}$.
(2) $q_{wealth} \approx \frac{45}{15} - 1 = 2$. (3) $q_{individual} \approx \frac{1}{2}(q_{wealth} + q_{tradition}) = 1.5$. (4) $q_{nation} \approx \frac{4}{3}0.288 = 0.384$. (5) $q_{global} \approx \frac{1}{2}\left(\frac{1.72}{16.76} + \frac{3}{5}\right) = 0.35$.

$c_1 \approx e^{0.889*0.7765} = 1.989$: (1) $u_{freedom} = 1 - \frac{-2+political\ rights+civil\ liberties}{18} = 0.889$, (2)
$u_{individual} \approx \frac{1}{2}(u_{Gini\ index} + u_{HDI}) = 0.5(0.539 + 0.301) = 0.42$. (3)
$u_{nation} = \frac{1}{3}(u_{democracy} + u_{debt} + u_{GDP}) = \frac{1}{3}(0.288 + 0.59 + 0.25) = 0.376$. (4) $u_{Global} \approx$
$u_{economical\ ties} = -0.0225$.

***Current situation:*** The wealth distribution in Brazil is among the worst in the world (the Gini index is 0.539), and in particular the wealth of the top one percent is the largest from all nations studied here (the top 1% owns about 45% of all wealth is Brazil). Brazil's basic relative noise is very small and follows, $c_1/\widetilde{T}_1 \approx 0.085$. Thus, Brazil should reduce $\widetilde{T}$, while shifting resources in favor of the poorest 80%.

***Study of the following nations will appear online at www.flomenbom.net:*** Study of additional countries will appear during the time. The first new nations will include: Hungary, South Korea, North Korea, India, South Africa, Australia, New Zealand, Argentina, Ivory Coast, and Yemen.

***References***

1. See the official website, http://www.freedomhouse.org, and the report at,

http://www.freedomhouse.org/template.cfm?page=21&year=2011

2. See the governmental website: http://www.census.gov/

3. http://www.beta.undp.org

4. The HDI report: http://hdr.undp.org/en/media/HDR_2010_EN_Table3_reprint.pdf

5. Gini, C (1909) Concentration and dependency ratios (in Italian). English translation in *Rivista di Politica Economica*, 87 (1997), 769-789.

6. About nations' GDP, debt, Gini index, and top 10% from the CIA:

https://www.cia.gov/library/publications/the-world-factbook/index.html

7. About nations' Gini index, top one percent & top 10 percent from the World Bank:

http://data.worldbank.org/indicator

8. M. Brenner, D. Galai, Fin. Ana. J., 61 (1989).

9. Please see the official website: http://www.gallup.com

10. http://www.economistgroup.com/; where the report at, Democracy index:

http://www.economist.com/media/pdf/DEMOCRACY_INDEX_2007_v3.pdf

11. About trading from WTO:

http://stat.wto.org/CountryProfile/WSDBCountryPFReporter.aspx?Language=E

12. About the top 1 percent and 1 in a thousand:

http://sociology.ucsc.edu/whorulesamerica/power/wealth.html

13. About the top 1 percent: http://www.vanityfair.com/society/features/2011/05/top-one-percent-201105

14. A. Chapnick, The Middle Power. *Canadian Foreign Policy* 7,2 (Winter 1999): 73-82.

15. About the IMF shares from the IMF:

http://www.imf.org/external/np/sec/memdir/members.aspx